\documentclass[showpacs,preprint,11pt,nofootinbib,superscriptaddress,citeautoscript,eqsecnum]{revtex4}
 \usepackage{amsmath,amsfonts,amssymb}
\usepackage{color}
\definecolor{darkblue}{rgb}{0.0,0.0,0.8}
\definecolor{darkred}{rgb}{1,0.5,0.7}
\definecolor{Brown}{cmyk}{0, 0.8, 1, 0.6}
\usepackage{setspace}
\usepackage{hyperref}
\hypersetup{colorlinks=true, citecolor=darkblue, linkcolor=Brown,  urlcolor=darkblue}
\newcommand{\nc}{\newcommand}
\nc{\ba}{\begin{eqnarray}}
\nc{\ea}{\end{eqnarray}}
\newcommand\be{\begin{equation}}
\newcommand\ee{\end{equation}}

\nc{\x}{{\bf{x}}}

\input{epsf.sty}

\begin{document}

 \title{dS solutions with co-dimension two branes in six dimensions}
\author{Hamid R. Afshar}
\email{afshar@hep.itp.tuwien.ac.at}
\affiliation{Institute for Theoretical Physics, Vienna University of Technology, Wiedner Hauptstrasse 8-10/136, A-1040 Vienna, Austria, Europe}

\author{Hassan Firouzjahi}
\email{firouz@ipm.ir}
\affiliation{School of Astronomy, Institute for Research in Fundamental Sciences (IPM), P. O. Box 19395-5531, Tehran, Iran}

\author{Shahrokh Parvizi}
\email{parvizi@modares.ac.ir}
\affiliation{Department of Physics, School of Sciences, Tarbiat Modares University, P.O. Box 14155-4838, Tehran, Iran}
 \date{\today}
 \preprint{TUW-12-35}

\begin{abstract}
We investigate phenomenological 4D dS solutions with co-dimension two branes and finite 4D Planck mass in
six dimensions. We present the conditions under which six-dimensional compactifications  with holomorphic axion-dilaton field or models with pure gravity with local sources can yield 4D dS solutions.  
Different classes of solutions are specified by a holomorphic function describing different embeddings of multiple conical branes.  Depending on the local singularities of this holomorphic function and the 
topology of the compact dimension one has to introduce 3-branes creating a deficit angle equal  to $\pi$ and/or 4-branes with positive tensions.
\end{abstract}

\maketitle
\tableofcontents
\section{Introduction}

Brane-world and compactification scenarios in extra dimensions \cite{ArkaniHamed:1998rs} 
are at the intersections of different fields of high energy physics.
It has been employed to study particle physics implications such as addressing the gauge hierarchy problem \cite{Randall:1999ee,Randall:1999vf}
and in the fundamental aspects of gravitational theories such as explaining  the cosmological constant problem  \cite{ArkaniHamed:2000eg,Kachru:2000hf},
for a review of arguments why extra dimensions is useful for cosmological constant problem see \cite{Burgess:2007ui}.
According to brane-world scenarios our observable universe is confined on a 3-brane in extra dimensions. For instance
Randall and Sundrum proposal is based on a five-dimensional gravity with a
negative bulk cosmological constant. Imposing suitable boundary conditions at the 3-branes positions the vacuum energy of
four-dimensional space-time is absorbed by the extra transverse dimension.
The fact that the vacuum energy of 3-branes in six dimensions contributes only to the energy momentum tensor in the transverse directions
has made the co-dimension two brane
world scenarios very attractive as a possible solution of the cosmological constant problem \cite{Rubakov:1983bz, Gherghetta:2000qi, Chen:2000at, Navarro:2003vw}.
These ideas have been considered in different contexts such as flux compactification \cite{Carroll:2003db, Aghababaie:2003wz, Navarro:2003bf, Cline:2003ak, Gibbons:2003di, Aghababaie:2003ar, Burgess:2004kd, Burgess:2004dh, Lee:2005az},
non-linear sigma models \cite{GellMann:1984mu,GellMann:1985if,Lee:2004vn,RandjbarDaemi:2004ni}
and specifically in axion-dilaton gravity   \cite{Kehagias:2004fb,Nair:2004yu,Kehagias:2005dp}. The construction of
co-dimension two branes as solutions of axion-dilaton gravity goes back to  string theory \cite{Greene:1989ya}.

Cosmological observations indicate that our universe is endowed with a very small positive cosmological constant which makes the study of de-Sitter (dS) compactifications very suggestive.
However there are no-go theorems which prohibits dS solutions in string theory or supergravity setups.
Maldacena and Nunez  (MN) \cite{Maldacena:2000mw}
showed that there is no non-singular warped dS compactification with finite effective Newton's constant, for a new revisit of the
problem see \cite{Douglas:2010rt}.
They also proved that it is not possible to find warped compactifications which have only singularities where
the warp factor is non-increasing as we approach the singularity. This rules out any smooth Randall-Sundrum-like compactification
of usual supergravity theories.
This no-go theorem is evaded in string theory by inclusion of localized sources with positive/negative tensions,
D-branes/O-planes \cite{Verlinde:1999fy,Giddings:2001yu,Kachru:2003aw,Burgess:2003ic,Giddings:2005ff,Burgess:2011rv,VanRiet:2011yc}. For a recent
overview of some open questions on de-Sitter physics see \cite{Anninos:2012qw}
and references therein.

In this paper we present general warped dS solutions of 6D pure gravity. Higher dimensional dependence of solution is encoded in an arbitrary
holomorphic function $f(z)$. By choosing different forms of this function one can introduce different forms of identifications and singularities in the two
extra dimensions. As an example we present three different warped compactifications;  (a) `co-dimension one compactification' which is the analogue of Randal-Sundrum II
picture in 6D with a positive tension 4-brane at the singularity, (b)
`co-dimesion two compactification' in which the 3-brane introduces a conical singularity of $\pi$ and (c) `compactification with double periodic functions'
which needs a 4-brane with smeared 3-branes wrapped around a cycle of torus. It also requires four 3-brane conical singularities.
We compute the matching conditions which measure the back-reactions of co-dimension two branes on the internal geometry by using complex variables,
for an alternative method see \cite{Burgess:2008yx,Bayntun:2009im}.

We emphasis that we follow a phenomenological approach, common to many brane-world scenarios.
The goal is to see whether or not it is possible to obtain a four-dimensional dS solution from a higher-dimensional gravitational theory with local sources behaving as $p$-branes.  Some ingredients of our set up may arise from string theory and supergravity, but in general we allow phenomenological setup which may not have string theory origins.

The paper is organized as follows. In section \ref{theory} we introduce our setup, find a general
formula for the Euler number of the internal manifold and review the well-known MN no-go theorem in obtaining a
dS solution for this setup.
In section \ref{AXIDIL} we map the problem into a six-dimensional gravity plus axion-dilaton system in terms of which the equations of
motion is presented.
We divide the solutions of the equations of motions into two categories; holomorphic and constant axion-dilaton field
and present them respectively in sections \ref{CASE1} and \ref{consttau}.  The former is the same as the type IIB solution obtained in \cite{Greene:1989ya, Bergshoeff:2006jj}.
Different solutions for pure gravity model are presented in  sections \ref{flat} and \ref{nonflat}.
The summary and discussions are given  in section \ref{summary}.
We relegate some technical issues regarding the contribution of branes  to the energy momentum tensor and the singularity analysis in complex planes into the Appendices.


\section{Action and field equations}\label{theory}
We consider five form flux $F_{(5)}=dC_{(4)}$ and dilaton, $\phi$ in 6D space-time. The action is
\ba \label{fullaction}
S=\frac{1}{2\kappa^2}\int d^6x
\sqrt{-g} \left( R -\partial_M\phi\partial^M\phi- \frac{e^{-2 \phi}}{5!} F_{(5)}^2
\right)+S_{\text{loc}}
\ea
where  $\kappa$ is the six-dimensional gravitational mass scale and $S_{\text{loc}}$ is the contribution of local D$p$-branes to the action
which is introduced in Appendix \ref{brane-T}. The motivation for this action originates from string theory and supergravity. However, as we mentioned before, we follow a phenomenological approach and also allow  setups which may not have string theory origins. In particular, we are mainly interested in the gravitational effects of the local sources, so the D$p$-branes  may not be charged under $C_{(4)}$ as required in string theory, corresponding to 
$\mu_{p}=0$. In this view our local sources may be called $p$-branes rather than the standard D-branes in string theory\footnote{One may try to derive action (2.1) by starting from a dilatonic background including $F_{(9)}$ in IIB SUGRA (F-theory) and
then compactify and dimensionally reduce it to six dimension such that only 5-form, dilaton and
graviton survive. One may then consider charged local sources in 6D to be originated from charged D-branes in 10-dimensional theory. Neutral $p$-branes in 6D can also be included as brane anti-brane superpositions of D-brane sources in ten-dimensional theory where they have neutral net charges consistent with setting form fields equal to zero. We don't intend to give a rigorous prescription for this dimensional reduction, since we are motivated by phenomenological purposes.}.

The equations of motion are
\ba\label{eom5}
R_{AB} &=&  \partial_A \phi \partial_B \phi-  \frac{e^{-2 \phi}}{5!}F^2g_{AB}+\frac{e^{-2 \phi}}{4!}F_{AB}^2+ \left(T_{AB}-\frac{1}{4}Tg_{AB}\right)_{\text{loc}}
 \\
\nabla^2 \phi &=& \frac{1}{\sqrt{-g}} \partial_M\left( \sqrt{-g}g^{MN} \partial_N \phi
\right) =  -\frac{1}{ 5!}e^{-2 \phi} F^2\label{eom5.1}\\
d(e^{-2\phi}* F)&=&\partial_A \left(  \sqrt{-g} e^{-2 \phi} F^{ABCDE}
 \right) =0
\ea
with the Bianchi identity,
\ba\label{bianchi}
dF&=&\partial_{[M}F_{NPQRS]}=0\,.
\ea

We are interested in the following warped ansatz,
\ba\label{ansatz2}
ds^2&=&e^{2w(y)}\tilde{g}_{\mu\nu}dx^{\mu}dx^{\nu}+\hat{g}_{mn}dy^mdy^n \, ,
\nonumber\\
F_{0123m}&=&Q(y)\epsilon_{0123m} \, ,\nonumber\\
\phi&=&\phi(y) \, .
\ea
In this notation the four-dimensional coordinates are denoted by the Greek indices $\{ x^\mu\}$ while the internal coordinates are denoted
by the Latin indices $\{ x^m\}$. Also $\tilde{g}_{\mu\nu}(x^\alpha)$ is the four-dimensional metric which we take
to be maximally symmetric. Later on we shall concentrate on the case in which
$\tilde{g}_{\mu\nu}$ has the dS form. Finally the internal metric, which only depends on
$\{ x^m\}$, is denoted by $\hat{g}_{mn}$.

The components of Ricci tensor are
\ba\label{Ricci}\label{eom14}
R_{\mu\nu}&=&\tilde R_{\mu\nu}(\tilde{g})-\frac{1}{4}e^{-2w}\hat\nabla_m\hat\nabla^me^{4w}\tilde{g}_{\mu\nu}\\
R_{mn}&=&\hat R_{mn}(\hat{g})-4e^{-w}\hat\nabla_m\hat\nabla_n e^{w}\label{eom12}
\ea
in which $\hat\nabla_m$ represents the covariant derivative with respect to the internal metric
$\hat{g}_{mn}$. For the two-dimensional compact manifold, $\hat R_{mn}(\hat{g})=\hat K(y)\hat{g}_{mn}$, where $\hat K(y)$ is the Gauss curvature of the manifold.
Assuming maximally symmetric four-dimensional space-time, implies
\ba
\tilde R_{\mu\nu}(\tilde{g}) =3\lambda\, \tilde{g}_{\mu\nu}
\ea
with $\lambda$ being a constant. For a dS solution, we take $\lambda >0$.
On the other hand from \eqref{eom5} and \eqref{Branecont1} we have \cite{Douglas:2010rt},
\ba\label{eom14.1}
g^{\mu\nu}R_{\mu\nu}&=& \kappa^{2}(p-3)\,T_p\delta(\Sigma)\\
g^{mn}R_{mn}&=&\frac{\kappa^{2}(7-p)}{2}\,T_p\delta(\Sigma)+e^{-2\phi}Q(y)^2\label{eom14.2}
\ea
where $T_p$ is brane tension which we take to be positive.  As a result,  using \eqref{eom5.1}
and \eqref{eom14}--\eqref{eom14.2}, the equations
for $\phi$, $R_{mn}$ and $R_{\mu\nu}$ give
\ba
\hat\nabla_{m}\left(e^{4w}\hat\nabla^m \phi\right)&=& e^{4w}e^{-2 \phi}Q(y)^2 \label{Rphi}\,\\
\hat K(y)&=&2e^{-w}\hat\nabla^2e^w+\frac{\kappa^2}{4} (7-p)\,T_p\delta(\Sigma)+ \frac{1}{2}e^{-2\phi}Q(y)^2\label{Rphi2}\, \\
12\lambda e^{2w}&=&\hat\nabla^2e^{4w}-\kappa^2 (3-p)\,T_p\,\delta(\Sigma) e^{2w}\label{Rphi1}\, .
\ea
In general if we assume that the internal manifold $Y$ is compact without boundary,
integrating on both sides of \eqref{Rphi} the left hand
side vanishes and we have
\ba\label{lambda}
\int_Ye^{4w}e^{-2 \phi} Q(y)^2=0 \, ,
\ea
which shows that $Q(y)$ should identically be zero since the integrand is positive-definite. There is an exception however, when
the warp factor is constant and the field strength is $Q(y)\sim\partial_y\phi \,e^{\phi}$, such that LHS and RHS of \eqref{Rphi}
is a total derivative, $\hat\nabla^2 e^{-\phi}=0$, altogether, we will discuss this case in section \ref{CASE1}.
\subsection{Euler character}
On the other hand, in order to have a compact two dimensional manifold, we need to have the Euler number
to be positive. With this asumption -- compactness -- we can also find the Euler character with partially integrating \eqref{Rphi2},
\ba\label{TrGauss}
\chi_{{}_E}&\equiv&\frac{1}{2\pi}\int_Y\sqrt{\hat g}\,\hat K(y) d^2y\nonumber\\
&=&\frac{1}{2\pi}\int_Yd^2y\sqrt{\hat g}\,\left(2(\hat\nabla w)^2+\frac{\kappa^2 (7-p)}{4}\,T_p\delta(\Sigma)+\frac{1}{2}e^{-2\phi}Q(y)^2\right) \, ,
\ea
which is positive definite and shows that a non-constant warp factor is consistent with compactification assumption. If the manifold has a boundary there will be an additional contribution to the Euler character,
\ba\label{Euler}
\delta\chi_{{}_E}=\frac{1}{2\pi}\int_{\partial Y} ds\left(2n\,\cdotp\partial w+ \hat{k}\right)
\ea
where $ds$ is the line element along the boundary and $\hat{k}=-t^an_b\hat\nabla_a t^b$ is the geodesic curvature of the boundary, with $t^a$ and $n_b$ unit
vectors tangent and outward normal to the boundary, respectively.


\subsection{No-go theorem for dS vacuum}
In the special case of constant warp factor, $w=0$, from Eq. (\ref{Rphi1}) to get a
dS solution we need $p > 3$.
Now the question is if one can
have warped dS space-time such that the conditions from Eqs. (\ref{Rphi})-(\ref{Rphi1}) are satisfied at the same time.
The Maldacena-Nunez no-go theorem \cite{Maldacena:2000mw} in our context is as follows. Assuming $\lambda \ge 0$,
from \eqref{Rphi1} we have
\ba\label{malda}
e^{4w}\hat\nabla^2 e^{4w}-e^{6w}\kappa^2 (3-p)T_p\delta(\Sigma) \geq0 \, .
\ea
If we integrate \eqref{malda} by parts we conclude that
\ba\label{no-go}
\int_Yd^2y\sqrt{\hat g}\,(\hat\nabla e^{4w})^2-\int_{\partial Y}ds\,e^{4w}n\,\cdotp\hat\nabla e^{4w}+\kappa^2 (3-p)T_p\int_Ye^{6w}\delta(\Sigma)\leq0 \, .
\ea
Assuming we have no source other than 3-branes, $p=3$, and the internal manifold is compact such that the boundary term disappears, then Eq. (\ref{no-go}) implies that the warp factor should be constant and $\lambda=0$. This is a manifestation of MN no-go theorem in our setup. To avoid MN no-go theorem in having a dS solution we need to have local branes with dimensions  $p>3$ or boundaries and non-constant warp factor.

\section{Axion--dilaton gravity}\label{AXIDIL}
From now on we only work with the bulk part of the action \eqref{fullaction} and take into account the sources when necessary. Moreover
for ease we introduce axion by using the Hodge duality
\ba
d\chi=e^{-2\phi} \ast F\,.
\ea
Inserting this into \eqref{fullaction},  we find the following action
\ba \label{fullaction1}
S=\frac{1}{2\kappa^2}\int d^6x
\sqrt{-g} \left( {R} -\partial_M\phi\partial^M\phi - e^{2 \phi} \partial_M\chi\partial^M\chi 
\right)\,,
\ea
%
and the corresponding equations become
\begin{gather}\label{eom6}
R_{AB} = 
\partial_A \phi \partial_B \phi+  e^{2 \phi}  \partial_A \chi \partial_B \chi
 \\
\nabla^2 \phi = \frac{1}{\sqrt{-g}} \partial_M\left( \sqrt{-g}g^{MN} \partial_N \phi
\right) = e^{2 \phi}g^{MN}\partial_M\chi\partial_N\chi\\
\partial_A \left(  \sqrt{-g} e^{2 \phi} g^{AB}\partial_B\chi
 \right) =0\label{eom9}\,.
\end{gather}
We can introduce the `axion-dilaton field' as a complex combination of the two real scalar fields by  $\tau=\chi+ie^{-\phi}$. The action can be rewritten as follows
\ba \label{fullaction2}
S=\frac{1}{2\kappa^2}\int d^6x
\sqrt{-g} \left( R -\frac{\partial_M\tau\partial^M\bar{\tau}}{(\text{Im}\, \tau)^2}
\right) \, ,
\ea
which is invariant under the following SL(2,$\mathbb{R}$) transformation,
\ba
 \tau\rightarrow\frac{a\tau+b}{c\tau+d}
\ea
where $(a,b,c,d)\in \mathbb{R}$ and $ad-bc=1$, while metric is held fixed.
The equations \eqref{eom6} -- \eqref{eom9} can be written as
\begin{gather}\label{eom13}
R_{MN}-\frac{1}{4\left(\text{Im}\, \tau\right)^2}(\partial_M\tau\partial_N\bar \tau +\partial_M\bar \tau\partial_N\tau)=0\\
\nabla^M\nabla_M\tau +\frac{i\nabla^M\tau\nabla_M\tau}{\text{Im}\,\tau}=0\label{KG1}\,.
\end{gather}
We find it very convenient to go to the complex $z$-plane in which $\ell z= y_1+ i y_2$ and
$\ell\bar z = y_1 - iy_2$ are dimensionless and $\bar z$ represents the complex conjugate of $z$. Furthermore, we can go to the conformal gauge in which  $\hat g_{mn}dy^m dy^n=\ell^2 e^{\Omega(z,\bar z)}dzd\bar z$  so our metric is given by
\ba
\label{metric-conformal}
ds^2&=&e^{2w(z, \bar z)}\tilde{g}_{\mu\nu}dx^{\mu}dx^{\nu}+ \ell^2 e^{\Omega(z,\bar z)}dzd\bar z\, ,
\ea
where $\ell$ has dimension of length and denotes the size of internal space.
With this ansatz, the components of Einstein tensor are
\ba
\label{G00}
\ell^{2}G_{\mu\nu}&=&\left(-3\tilde{\lambda}+\left(12\partial\bar\partial w+24\partial w\bar \partial w +2 \partial\bar\partial \Omega \right) e^{2w-\Omega} \right) \tilde g_{\mu\nu}\\
\label{Gz-barz}
\ell^{2}G_{z\bar z}&=&-3\tilde{\lambda}\, e^{-2w+\Omega}+4\partial\bar\partial w+ 16\partial w\bar\partial w\\
\label{Gzz}
\ell^{2}G_{zz}&=&-4 \partial^2 w -4 (\partial w)^2 +4\partial w\partial \Omega
\ea
where $\partial\equiv\partial_z\,$, $\bar\partial\equiv\partial_{\bar z}$ and $\tilde{\lambda}=\lambda \ell^2$ is dimensionless.
Therefore, our field equations, Eqs. \eqref{eom13}--\eqref{KG1}, can be written as
\ba
3\tilde{\lambda}-4(\partial\bar\partial w+4\partial w\bar\partial w)e^{2w-\Omega}&=&0\label{eom21b}\\
\partial^2w+(\partial w)^2-\partial w\partial \Omega+\frac{1}{8 (\text{Im}\,\tau)^2}\partial \tau \partial \bar \tau &=& 0
\label{eom36b}\\
\partial\bar\partial\Omega +4(\partial w\bar \partial w+\partial\bar\partial w)+
\frac{1}{4(\text{Im}\,\tau)^2}(\partial\tau\bar\partial\bar\tau +\partial\bar\tau\bar\partial\tau)
&=&0
\label{eom19b}\\
\partial \bar \partial \tau + \frac{i}{\text{Im}\,\tau} \partial \tau \bar \partial \tau +
2 (\partial w \bar \partial \tau + \bar \partial w  \partial \tau) &=&0\label{eom18bb}
\ea
Before presenting the solutions for this set of equations in some cases, we should comment on singularities and sources. The delta function singularities corresponding to brane sources arise from Laplacians $\partial\bar\partial w$ and $\partial\bar\partial\Omega$ in \eqref{G00} and \eqref{Gz-barz}. We call  singularities which originate from $\partial\bar\partial w$ and $\partial\bar\partial\Omega$ the warp-type and the conformal-type singularities, respectively. The warp-type singularity appears in both $\mu\nu$ and $z\bar z$ directions in \eqref{G00} and \eqref{Gz-barz} so it corresponds to 4-branes (and possible smeared 3-branes) singularity. On the other hand the conformal-type singularity corresponds to isolated 3-branes.

\subsection{Holomorphic axion-dilaton field}\label{CASE1}
One interesting example is when $\tau$ is holomorphic,
\ba\label{holomor}
\bar \partial \tau = \partial \bar \tau = \bar \partial \partial \tau =0\,.
\ea
In this case, Eq. (\ref{eom18bb}) results in
\ba
\label{cond1}
\bar \partial w \partial \tau =0 \, .
\ea
This equation can be satisfied by either $\bar \partial w=0$ or $ \partial \tau=0$. The latter together with \eqref{holomor} gives $\tau=$ constant which we will study in next section.

Now consider the first case, in which $\bar \partial w=0$, indicating that $w$ is holomorphic too. Plugging this in Eq. (\ref{eom21b}) implies that $\lambda=0$.
This is an interesting result: for a holomorphic and non-constant $\tau$, $\partial \tau\neq 0$, we should require $\lambda=0$.

Using the following relation (valid for $\partial w =0$)
\ba\label{identity}
\partial \bar \partial \ln \text{Im}\,\tau  =  \frac{-1}{4 (\text{Im}\,\tau)^2} (  \partial \tau \bar \partial \bar \tau + \partial \bar \tau \bar \partial \tau )\,,
\ea
from  Eq. (\ref{eom19b}) one obtains
\ba\label{holomorphic}
\bar \partial \partial (\Omega - \ln \text{Im}\,\tau ) =0 \, .
\ea
This can be solved to obtain
\ba
\label{hol-ans1}
\Omega = - \phi + h(z) + \bar h(z) \, ,
\ea
in which $h(z)$ is an arbitrary holomorphic function. Then the metric reads as
\be\label{flat4D}
ds^2=\eta_{\mu\nu}dx^\mu dx^\nu+\ell^2\frac{\text{Im}\,\tau}{|f(z)|^{2}}  dz d\bar z
\ee
where $f(z)=e^{-h(z)}$. This solution resembles the well known F-theory compactification of type IIB supergravity to eight dimensions,
with space filling  7-branes.  The inclusion of  branes and SL(2,$\mathbb{Z}$) properties of this set up were discussed in details in \cite{Greene:1989ya, Bergshoeff:2006jj}.

\subsection{Constant axion-dilaton field}\label{consttau}
Now  consider the other solution of Eq. (\ref{cond1}) in which $\partial \tau =0$, so $\tau$ is a constant. This is pure gravity without any matter. Then equations \eqref{eom21b} - \eqref{eom18bb} simplify to
\ba
3\tilde{\lambda}-4(\partial\bar\partial w+4\partial w\bar\partial w)e^{2w-\Omega}&=&0\label{eom21bb}\\
\partial^2w+(\partial w)^2-\partial w\partial \Omega &=& 0
\label{eom36bb}\\
\partial\bar\partial\Omega +4(\partial w\bar \partial w+\partial\bar\partial w)
&=&0
\label{eom19bb}
\ea
Equation \eqref{eom36bb} is easily solved
\ba
{\bar\partial e^w= f( z)e^{\Omega}}\label{eom33}
\ea
where $f$ is an arbitrary holomorphic complex function.
Plugging \eqref{eom33} into \eqref{eom21bb} and \eqref{eom19bb} we find,
\begin{gather}
\partial e^w=\bar fe^{\Omega}\label{eom28}\\
\bar\partial(\Omega +3w+\ln f(\bar z))=\frac{3\tilde{\lambda}}{4}\frac{e^{-w}}{\bar f}\label{eom29}\\
\partial\bar\partial( \Omega +3w)=\frac{3\tilde{\lambda}}{4}\frac{\partial e^{-w}}{\bar f}\, ,\label{eom30}
\end{gather}
provided that $f(\bar z)\neq0$. However, it is important to note that this condition can be violated at finite number of points corresponding to the position of local branes.
These equations gives
\ba\label{eom10}
\bar\partial(4w+\ln \partial w)=\frac{3\tilde{\lambda}}{4}\frac{e^{-w}}{\bar f}\\
\partial(4w+\ln \bar\partial w)=\frac{3\tilde{\lambda}}{4}\frac{e^{-w}}{ f}\,.
\ea
These two equations further can be written as
\ba\label{eom20}
 W^3\bar\partial W-\frac{W^3}{\bar f}&=&d(\bar z)\nonumber\\
 W^3\partial W-\frac{W^3}{ f}&=&d(z)
\ea
where $d(z)$ is an integration function and
\ba
W(z,\bar z) \equiv \frac{4}{\tilde{\lambda}} e^{w(z,\bar z)}.
\ea
The conformal factor in this case becomes
\ba\label{eom303}
 e^\Omega=\frac{\partial(e^w)}{\bar f}=\frac{\tilde{\lambda}}{4|f(z)|^2}\left(1+\frac{d(\bar z) f(\bar z)}{W^3} \right)\nonumber\\
 e^\Omega=\frac{\bar \partial(e^w)}{f}=\frac{\tilde{\lambda}}{4|f(z)|^2}\left(1+\frac{d(z) f(z)}{W^3}\right) \, .
\ea
The condition that $\Omega$ is real  enforces us to take $d(z)\propto 1/f(z)$, so we have
\ba\label{eom121}
{
\partial W=\frac{1}{f(z)}\left(1+\frac{\sigma^3}{W^3}\right)\quad\text{and}\quad \bar \partial W=\frac{1}{ f(\bar z)}\left(1+\frac{\sigma^3}{W^3}\right)}
\ea
where $\sigma$ is a real proportionality constant which physically turns out to be $\sigma=0,\pm1$.  As a result,
\ba
{
e^\Omega=\frac{\tilde{\lambda}}{4|f(z)|^2}\left(1+\frac{\sigma^3}{W^3}\right).}
\ea
So the line element is
\ba
\label{metric-final}
ds^2=\frac{\tilde{\lambda}^2}{16}W^2\tilde g^{\text{(A)dS}}_{\mu\nu}dx^\mu dx^\nu
+\ell^2\frac{\tilde{\lambda}}{4|f(z)|^2}\left(1+\frac{\sigma^3}{W^3}\right) \,dzd\bar z \, .
\ea
When $\sigma=0$ we find $e^\Omega=\frac{\tilde{\lambda}}{4|f(z)|^2}$, thus $\lambda$ should be necessarily positive.  For $\sigma\neq 0$, $\lambda$ can be either positive or negative depending on the sign of $1+\sigma^3/W^3$. In $\sigma\neq 0$ case, by an appropriate scaling of parameters, we can set $\sigma=1$ without loss of generality. So we have two distinguished cases $\sigma= 0$ and $\sigma=1$.

The $4$D-Planck mass is obtained by looking at the 4D effective action,
\ba
S_{6}= \kappa^{-2} \int d^2z \sqrt{\hat g} e^{2w}\int d^4x\sqrt{-\tilde g}\tilde R^{(4)} \, .
\ea
Therefore the 4D-Planck mass $M_P$ can be derived as
\ba\label{eom1111}
M_{P}^2&=& \kappa^{-2}\int d^2z\sqrt{\hat g} \,e^{2w}\nonumber\\
&=&\frac{\tilde{\lambda}^3 \ell^2}{64 \kappa^{2}}\int_R d^2z \frac{W^2}{|f(z)|^2}\left(1+\frac{\sigma^3}{W^3}\right)\,,
\ea
which should be finite. This depends on the function $f(z)$ as a free holomorphic function and also the domain of integration $R$.
The warp factor $W(z,\bar z)$ should be obtained from the equation \eqref{eom121}. In the following sections we solve this equation when $\sigma=0$ and $\sigma =1$ separately.

\section{Locally flat solutions ($\boldsymbol{\sigma=0}$)}
\label{flat}

By setting $\sigma=0$ in \eqref{eom121} the metric takes the following simple form
\ba
\label{metric-sigma0}
ds^2=\frac{\tilde{\lambda}^2}{16}W^2\tilde g^{\text{dS}}_{\mu\nu}dx^\mu dx^\nu
+\frac{\tilde{\lambda} \ell^2}{4|f(z)|^2}\,dzd\bar z
\ea
where the warp factor is simply solved from equations (\ref{eom121}),
\be\label{Wf=0}
W(z,\bar z)=\xi(z)+\bar\xi(\bar z)+c_1\,\qquad\text{with}\qquad\xi(z)=\int^z_0\frac{dt}{f(t)}\,.
\ee
where $c_1$ is an integration constant. As a result the $z$-plane is mapped to the $\xi$-plane by a conformal transformation $f(z)=dz/d\xi$.
Therefore, independent of the form of $f(z)$, all solutions are locally the same:
\ba\label{locallC}\label{eomff}
ds^2=\frac{\tilde{\lambda}^2}{16}(\xi+\bar\xi+c_1)^2\tilde g^{\text{dS}}_{\mu\nu}dx^\mu dx^\nu +\frac{\tilde{\lambda} \ell^2}{4}\, d\xi d\bar\xi\,.
\ea
One can also calculate the following scalars for \eqref{locallC},
\be
R=0,\qquad R_{\mu\nu\rho\sigma}R^{\mu\nu\rho\sigma}=0,\qquad C_{\mu\nu\rho\sigma}C^{\mu\nu\rho\sigma}=0\,,
\ee
where $R$ is the Ricci scalar, $R_{\mu\nu\rho\sigma}$ is the Riemann tensor and
$C_{\mu\nu\rho\sigma}$ is the Weyl tensor. The two-dimensional manifold is locally $\mathbb{C}$ for all $f$'s.
There are only topological degrees of freedom by inclusion of local singularities and identifications.
Depending on the type of singularity we consider 
separate parameterization for the complex plane in the following.

\subsection{Co-dimension one singularity}
In order to see how to deal with co-dimension one brane singulairties we present an example. Let us try the following function:
\ba
f(z)=-e^z\qquad\text{with}\qquad z=|y_1|+iy_2
\ea
in which we have identified $y_1$ and $-y_1$. Then from \eqref{Wf=0} we have (taking $c_1=0$),
\ba
W(z,\bar z) = ( e^{-z}+e^{-\bar z})= 2 e^{-|y_1|}\cos({y_2})
\ea
and the metric \eqref{metric-sigma0} reads as
\ba\label{cylindercase}
ds^2=\frac{\tilde{\lambda}^2  }{4 }e^{-2 |y_1|}\cos^2({y_2})\left(\tilde g^{\text{dS}}_{\mu\nu}dx^\mu dx^\nu\right) +\frac{\tilde{\lambda} \ell^2}{4 }e^{-2 |y_1|}\left(dy_1^2+dy_2^2 \right)\,.
\ea
We can interpret this solution topologically as a cylinder\footnote{In fact this choice of $f(z)$ maps $\mathbb{C}$ to the cylinder.} with $y_1$ and $y_2$ as longitudinal
and azimuthal coordinates, respectively and the identification $y_1$ and $-y_1$ as in Randall-Sundrum.
The 4D-Planck mass, taking into account the factor 2 from $\mathbb{Z}_2$ symmetry, is   
\ba
M_{P}^2&=& \frac{\tilde{\lambda}^3 \ell^2 }{8 \kappa^2 }\int_0^\infty dy_1 e^{-4 y_1}\int_0^{2\pi } dy_2 \cos^2({y_2})  \nonumber\\
\label{Mplanckcyl} &=& \frac{\tilde{\lambda}^3 \ell^2 \pi}{32 \kappa^2 }
\ea
which is finite.
To satisfy the junction conditions we look at Einstein equations,
\be
\label{Ein-eq}
G_{MN}\equiv R_{MN}-\frac{1}{2}R\,g_{MN}=\kappa^2 T_{MN}
\ee
in which $T_{MN}$, the contribution from the local source $p$-branes are summarized in
Appendix  \ref{brane-T}. Also the components of the Einstein tensor are given in Eqs. (\ref{G00})- (\ref{Gzz}).
Furthermore, for the case at hand  we have,
\be
e^{2w}=\frac{\tilde{\lambda}^2 }{4 }e^{-2 |y_1|}\cos^2({y_2})\,,\quad\text{and}\quad e^\Omega=\frac{\tilde{\lambda} }{4 }e^{-2 |y_1|}
\ee
which gives
\ba
\label{eom101}
G_{\mu\nu}&=&-8\lambda\cos^2y_2\,\delta(y_1) \,\tilde g_{\mu\nu}\\
G_{y_1y_1}&=&0\\
\label{eomy2y2}
G_{y_2 y_2}&=&-{8}\,\delta(y_1)
\ea
where $\delta(y_1)=\frac{1}{2}\frac{\partial^2|y_1|}{\partial{y_1^2}}\big|_{y_1=0}$.

Equations \eqref{eom101} and (\ref{eomy2y2}) introduce a 4-brane at $y_1=0$ with the tension given by
Eq. \eqref{Branecont}
\ba
T^a_b&=& - \frac{2T_4}{\ell\sqrt{ \tilde{\lambda}}}  \delta^a_b \delta (y_1)
\quad , \quad \{a, b \} =\{x^\mu, y_2\}
\ea
Matching the singular terms in Einstein equations  (\ref{Ein-eq}) yields the following results for the 4-brane tension
\ba
\label{T4}
T_4 = \frac{16}{\kappa^2 \ell^2  \sqrt{\lambda}} \, .
\ea
Interestingly, the 4-brane tension is positive and scales with $\frac{1}{\sqrt{\lambda}}$.

\subsection{Co-dimension two singularity}\label{co-dim2}
Here we consider the situation in which the internal manifold has the disc configuration.
In particular, we are interested to see if we can have dS solution with pure 3-branes localized on singular points inside the disc.
In order to identify the local singularities associated with the positions of 3-branes, we plug in the metric \eqref{metric-sigma0} into
the Einstein tensor and keep all potentially singular terms\footnote{We have used the fact that $\partial\bar\partial W=\frac{1}{2}(\partial\bar\partial+\bar\partial\partial)W$.}
\ba\label{singularterm}
G^\mu_{\nu}&=&\frac{1}{\tilde{\lambda}\ell^2}\left[24 W^{-1}f\bar f(\partial\frac{1}{\bar f}+\bar\partial\frac{1}{f})-8 f\bar f\partial\bar\partial \ln f\bar f \right]\delta^\mu_\nu\\
G_{z\bar z}&=&2W^{-1}\left(\partial\frac{1}{\bar f}+\bar\partial\frac{1}{f}\right)\\
G_{zz}&=&-\frac{4W^{-1}}{f}\,\partial\ln \bar f\\
G_{\bar z\bar z}&=&-\frac{4W^{-1}}{\bar f}\,\bar \partial\ln  f \, .
\ea
which is identically zero iff $f(z)$ is a regular function globally. However, since $f(z)$ is not globally regular, we have localized singularities corresponding to the position of branes.

Correspondingly, the $G_{rr}$, $G_{\theta\theta}$ and $G_{r\theta}$ components become
\ba\label{eomin2}
G_{rr}&=&-\frac{4W^{-1}e^{2i\theta}}{f}\partial\ln \bar f-\frac{4W^{-1}e^{-2i\theta}}{\bar f}\bar \partial\ln  f
+ \frac{4}{W} \left(\partial\frac{1}{\bar f}+\bar\partial\frac{1}{f}\right)\\
G_{\theta\theta}&=&\frac{4W^{-1}z^2}{f}\partial\ln \bar f+\frac{4W^{-1}\bar z^2}{\bar f}\bar \partial\ln  f
+ \frac{4\,r^2}{W} \left(\partial\frac{1}{\bar f}+\bar\partial\frac{1}{f}\right)\\
G_{r\theta}&=&-ir\frac{4W^{-1}e^{2i\theta}}{f}\partial\ln \bar f+ir\frac{4W^{-1}e^{-2i\theta}}{\bar f}\bar \partial\ln  f \, .\label{eomin2.2}
\ea
We are not interested in any type of singularity in $G_{r\theta}$, so we can assume $\partial\ln \bar f=\bar \partial\ln  f=0$. 
These conditions disallow functions like $f(z)=e^{1/z}$ with essential singulairty at $z=0$.
With these assumptions, \eqref{eomin2}-\eqref{eomin2.2} simplify to
\ba\label{eomin3}
G_{\theta\theta}=r^2G_{rr}=\frac{4\,r^2}{W} \left(\partial\frac{1}{\bar f}+\bar\partial\frac{1}{f}\right)\quad\text{and}\quad G_{r\theta}=0\,,
\ea
so in order to have a pure 3-brane solution which gives the singularity only in the $\{ x^\mu \}$ directions we should choose $f(z)$ such that \eqref{eomin3} is zero
globally but at the same time
the remaining contribution to $G^\mu_\nu=-\frac{8}{\tilde{\lambda}\ell^2}f\bar f\partial\bar\partial \ln f\bar f \delta^\mu_\nu$ gives a delta function. This corresponds to having only conformal-type singularities as introduced below \eqref{eom19b}.
Depending on the ansatz for $f(z)$ we find different configurations, where we present the elemntary example here.



\subsubsection*{Solutions with radial isometry}
\label{Match}

Let us try the following function
\be\label{fzalpha}
f(z)={z}^{1/\alpha}\qquad\text{with}\qquad z=re^{i\theta}
\ee
for which  we have,
\ba
W=\int\frac{dz}{f(z)}+\int\frac{d\bar z}{\bar f(\bar z)}&=&\frac{z^\beta+ \bar z^\beta}{\beta}=\frac{2}{\beta}\, r^\beta\cos \beta\theta\quad\qquad\text{}\quad \alpha\neq1\\
&=&\ln z+\ln \bar z=\ln (r^2)\qquad\text{}\quad \alpha=1\,
\ea
with $\beta=-1/\alpha+1=s+1$. Now we consider the cases $\alpha \neq 1$ and $\alpha=1$ separately.

Suppose $\alpha \neq 1$.
The metric \eqref{metric-sigma0} takes the following form
\ba\label{conesol}
ds^2=\lambda\rho^2\cos^2\beta\theta\left(\tilde g^{\text{dS}}_{\mu\nu}dx^\mu dx^\nu\right) +d\rho^2+\beta^2\rho^2d\theta^2    \quad \quad   (\alpha \neq 1) \quad \quad
\ea
where $\rho=\frac{\sqrt{\tilde{\lambda}}  \ell}{2 } \frac{r^{\beta}}{\beta}$.

To have a single-valued metric component, $\beta$ should be a multiple of $1/2$. Restricting to $0<\theta<2\pi \beta $ introduces two boundaries at $0$ and $2\pi \beta$. If we require periodic boundary condition by identifying these two boundaries, $\beta$ will be an integer or half integer.

When the periodicity of $ \theta$ is $2\pi$, this solution introduces a conical singularity at $r=0$ with the deficit angle $\Delta\phi=\frac{2\pi}{\alpha}$.
This represents a 3-brane located at $\rho=0$. This choice of $f(z)$ corresponds to the identification $z\sim e^{2\pi i\beta}z$; for $\beta=\frac{1}{N}$, this is just the quotient space $\mathbb{C}/\mathbb{Z}_N$.

The details of the singularity analysis, corresponding to the position of local 3-branes, are given in Appendix \ref{singular}. Using Eqs. \eqref{singularterm} and \eqref{delta1} with $s=-1/\alpha$, the contribution to the Einstein tensor is:
\ba
G^\mu_{\nu}&=&\frac{24}{ \ell^2\tilde{\lambda}}r^{2/\alpha}\left[\frac{\sin 2\pi\beta}{r^\beta\cos \beta\theta}-\frac{4\pi}{3\alpha} \right]\label{GMUNU}
\delta^2(z,\bar z)\,\delta^\mu_\nu\\
G_{\theta\theta}&=&r^2G_{rr}=4\, r^{2-\beta}\,\frac{\sin 2\pi\beta}{\cos \beta\theta}\,\delta^2(z,\bar z)\,.
\ea
Because $\beta$ should be positive ($s>-1$), the first term in \eqref{GMUNU} is dominant near the singularity $r\rightarrow0$.
In order to get rid of the singularity in $G_{rr}$ and $G_{\theta\theta}$  and the first term in \eqref{GMUNU} at the same time, 
the choices for non-integer $\beta>0$ are \eqref{svalues},
\begin{gather}
 \beta=\frac{1}{2},\frac{3}{2},\frac{5}{2},\cdots\,,
\end{gather}
corresponding to $\alpha=\pm2,-2/3,-2/5,\cdots$. These points corresponds to 3-branes singularity
with tension \eqref{Branecont},
\ba\label{T3-alpha}
T_3 = \frac{2 \pi}{\kappa^2\alpha}
\ea
However only for $\alpha=2$ ($\beta=1/2$) the deficit angle in \eqref{conesol} is physical (less than $2\pi$)
and leads to a positive 3-brane tension.

In order to find a finite Planck mass we should cut the geometry and identify outside
the disk with its inside. Starting with \eqref{conesol} we define the new coordinate $d u = - d\rho/\rho$ with the solutions $\rho=\rho_0 e^{-u}$ in which $\rho_0$ is the physical radius of the boundary 4-brane located at $u=0$. Imposing the $\mathbb{Z}_2$ symmetry $u\leftrightarrow-u$ the line element \eqref{conesol} transforms into
\ba
\label{new-cyl}
ds^2 = \lambda \rho_0^2 e^{-2 |u|} \cos(\beta \theta)^2  \left(\tilde g^{\text{dS}}_{\mu\nu}dx^\mu dx^\nu\right) + \rho_0^2 e^{-2 |u|} ( du^2 + \beta^2 d \theta^2)
\ea
which has the same form as  \eqref{cylindercase}.  This  leads to the following 4-brane tension
\be \label{tension2}
T_4=\frac{8}{\kappa^2\rho_0}\,.
\ee
The transformation of  \eqref{conesol} into  \eqref{new-cyl} or  \eqref{cylindercase}
is interesting. As mentioned before it is an example of the local equivalence of different choices of function $f$. Both solutions have a 4-brane as a boundary. The cross section of 4-brane with the $z$-plane is a circle with radius $\rho_0$ (for the cylinder $\rho_0=\sqrt{\tilde{\lambda}} \ell /2$), where \eqref{conesol} is the disc inside with a 3-brane sitting at the center and  \eqref{new-cyl} is the outside region of the circle. Since both are equipped with an inversion w.r.t. the circle, we conclude that both are equivalent with 3-brane mapped to infinity in  \eqref{new-cyl} solution. Also the 4-brane tensions \eqref{tension2} and \eqref{T4} are the same with $\rho_0=\sqrt{\tilde{\lambda}} \ell  /2$.

The 4D-Planck mass is (taking into account the factor 2 from $\mathbb{Z}_2$ symmetry)
\ba
M_{P}^2&=&2 \lambda\beta \kappa^{-2} \int_0^{\rho_0} d\rho \,\rho^3\int_0^{2\pi} d\theta \cos^2{\beta \theta}  \nonumber\\
&=& \frac{\lambda}{8\kappa^2 }\rho_0^4\left(\sin 4\pi\beta+4\pi\beta\right)
\ea
which the first term is zero. Taking $\beta=1/2$, as mentioned before,  and $\rho_0=\sqrt{\tilde{\lambda}} \ell/2$ one finds
\ba
M_{P}^2&=& \frac{\tilde{\lambda}^3 \ell^2 \pi}{64 \kappa^2}
\ea
which is one half of the \eqref{Mplanckcyl}. This is consistent, since $y_2$ coordinate is double cover of $\theta$ coordinate for $\beta=1/2$.

Now consider the case in which $\alpha=1$. The warp factor in this case depends only on $r$.
The line element \eqref{metric-sigma0} becomes
\be
ds^2=\frac{\tilde{\lambda}^2}{4}\rho^2\tilde g^{\text{dS}}_{\mu\nu}dx^\mu dx^\nu +\frac{\tilde{\lambda} \ell^2}{4}\left(d\rho^2+d\theta^2\right)  \quad \quad (\alpha=1) \quad \quad
\ee
where $\rho=\ln (r/r_0)$ is the physical distance and determines the hesitance near the singularity $r=r_0+\epsilon$,
\ba
L\sim\int\frac{d\epsilon}{r_0+\epsilon}\sim\ln {r_0}.
\ea
The circumference is always finite $L=\frac{\sqrt{\tilde{\lambda}} \ell\pi}{4}$.
The 4D-Planck mass can be read from \eqref{eom1111}
\ba
M_{P}^2
&\sim&\frac{\tilde{\lambda}^3 \ell^2}{\kappa^2 }\left[\int_\delta^{r_0-\epsilon} \frac{dr}{r}\, \left(\ln r/r_0\right)^2+\int_{r_0+\epsilon}^R \frac{dr}{r}\, \left(\ln r/r_0\right)^2\right] \nonumber\\
&\sim&\left(\ln r/r_0\right)^3|_\delta^{r_0-\epsilon}+\left(\ln r/r_0\right)^3|_{r_0+\epsilon}^R
\ea
which introduces an IR 4-brane at $r=\delta$ and a UV 4-brane at $r=R$.

The singularity contribution to the Einstein tensor is \eqref{delta}:
\ba
G^\mu_{\nu}&=&\frac{48}{\tilde{\lambda} \ell^2}\left[\frac{r^{2}}{\ln r^2}-\frac{r^{2}}{3} \right]2\pi\delta^2(z,\bar z)\,\delta^\mu_\nu\\
G_{\theta\theta}&=&r^2G_{rr}=\frac{8r^2}{\ln r^2}2\pi\,\delta^2(z,\bar z).
\ea
Note that the contribution to $G^\mu_\nu$ near the singularity $r\rightarrow0$ is like,
\be
G^\mu_{\nu}=-\frac{32\pi }{\tilde{\lambda} \ell^2}r^{2}
\delta^2(z,\bar z)\,\delta^\mu_\nu\,,
\ee
nevertheless the corresponding term in $G_{\theta\theta}$ remains intact.

This type of singularity can be generalized to $N$ 3-branes by choosing the following form of function $f$:
\be\label{multiple}
f(z)=\prod_{i=1}^{N}(z-z_i)^{1/\alpha_i}
\ee
where $z_i$ is the location of 3-branes. In the next subsection we introduce such an example.

\subsection{Double periodic solutions}\label{toroidal}

Let us choose the function $f$ as,
\be \label{fkz}
f(z)=\sqrt{(1-z^2)(1-k^2z^2)}\, .
\ee
This can be considered as a multiple-branch point generalization of \eqref{fzalpha} given in \eqref{multiple} with $N=4$ and $\alpha_i=2$, and is interesting for its global properties.

Using \eqref{fkz}, we have,
\be
\xi(z)=\int_0^z\frac{dt}{\sqrt{(1-t^2)(1-k^2t^2)}}\equiv \text{sn}^{-1}z\qquad\qquad|k|<1\,.
\ee
This is an \emph {elliptic integral of the first kind}. The inverse function of elliptic integral
\be
z=\text{sn}\, \xi
\ee
is called \emph{Jacobian elliptic function}.

There are two possible configurations for branch cuts as shown by the thick lines in the following pictures including typical non-trivial cycles:

\vspace{4em}
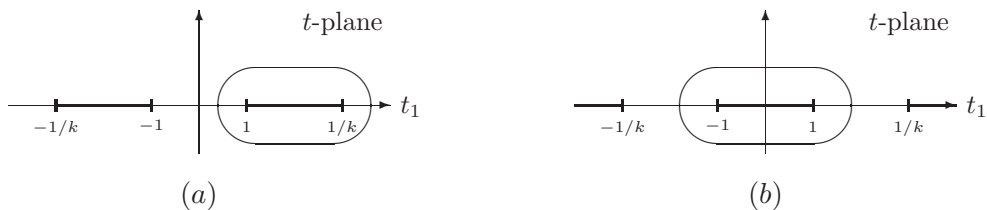
\begin{minipage}[ht]{0.45\textwidth}
\begin{center}
\setlength{\unitlength}{.01in}
\begin{picture}(200,50)
\linethickness{0.075mm}
\put(0,25){\vector(1,0){200}}
\put(100,0){\vector(0,1){75}}


\put(150,25){\oval(80,40)}

\thicklines

\put(25,22){\line(0,1){6}}
\put(75,22){\line(0,1){6}}
\put(125,22){\line(0,1){6}}
\put(175,22){\line(0,1){6}}

\put(25,25){\line(1,0){50}}
\put(125,25){\line(1,0){50}}

\put(25,9){\makebox(0,0)[b]{\tiny{$-1/k$}}}
\put(75,11){\makebox(0,0)[b]{\tiny{$-1$}}}
\put(125,11){\makebox(0,0)[b]{\tiny{$1$}}}
\put(175,9){\makebox(0,0)[b]{\tiny{$1/k$}}}
\put(100,-30){\makebox(0,0)[b]{$(a)$}}

 \put(211,19){\makebox(0,0)[b]{\small{$t_1$}}}
 \put(175,68){\makebox(0,0){\small{$t$-plane}}}
\end{picture}
 \end{center}

\end{minipage}
\begin{minipage}[ht]{0.45\textwidth}
\begin{center}
\setlength{\unitlength}{.01in}
\begin{picture}(200,50)
\linethickness{0.075mm}
\put(100,0){\vector(0,1){75}}
\put(0,25){\vector(1,0){200}}

%
%

\put(100,25){\oval(90,40)}

\thicklines

\put(25,22){\line(0,1){6}}
\put(75,22){\line(0,1){6}}
\put(125,22){\line(0,1){6}}
\put(175,22){\line(0,1){6}}

\put(0,25){\line(1,0){25}}
\put(75,25){\line(1,0){50}}
\put(175,25){\line(1,0){25}}

\put(25,9){\makebox(0,0)[b]{\tiny{$-1/k$}}}
\put(75,11){\makebox(0,0)[b]{\tiny{$-1$}}}
\put(125,11){\makebox(0,0)[b]{\tiny{$1$}}}
\put(175,9){\makebox(0,0)[b]{\tiny{$1/k$}}}
\put(100,-30){\makebox(0,0)[b]{$(b)$}}

 \put(211,19){\makebox(0,0)[b]{\small{$t_1$}}}
 \put(175,68){\makebox(0,0){\small{$t$-plane}}}
\end{picture}
 \end{center}

\end{minipage}
\vspace{2em}\\
If we define
\be
t-1=r_1 e^{i\theta_1},\qquad t+1=r_2 e^{i\theta_2},\qquad k t-1=r_3 e^{i\theta_3}\quad\text{and}\quad k t+1=r_4 e^{i\theta_4}
\ee
in the left figure we have
\be
0\leq\theta_1, \theta_4<2\pi\qquad\text{and}\qquad-\pi\leq\theta_2, \theta_3<\pi
\ee
while in the right figure
\be
0\leq\theta_2, \theta_3<2\pi\qquad\text{and}\qquad-\pi\leq\theta_1, \theta_4<\pi \, .
\ee
In both cases there is a relative minus sign once we cross the branch cut.

Note that $\xi(z)$ is not a single valued function of $z$, since the elliptic integral depends on the chosen contour on $t$-plane. Indeed a deformation of integration contour
leaves $\xi$ unchanged unless one goes around (or cross) a branch cut. This can be explained by the two well-known periods of the elliptic function. The first one is
\be
4K=4 \int_0^1\frac{dt_1}{\sqrt{(1-t_1^2)(1-k^2t_1^2)}} \, .
\ee
This corresponds to circling the $[-1,1]$ branch cut in the picture $(b)$ above.
The second period is introduced as
\be
K'= \int_0^1\frac{dt_1}{\sqrt{(1-t_1^2)(1-k'^2t_1^2)}}
\ee
where $k'^2=1-k^2$. By suitable change of variable, it can also be written as
\be
2i K'=2 \int_1^{1/k}\frac{dt_1}{\sqrt{(1-t_1^2)(1-k^2t_1^2)}} \, .
\ee
This is a period over the circle around $[1,1/k]$ branch cut in the picture $(a)$ above. Thus, for a fixed $z$, there are many $\xi$ for which
\be\label{periodicity}
\text{sn}\,(\xi +4K)=\text{sn}\,\xi\qquad\text{and}\qquad\text{sn}\,(\xi +2iK')=\text{sn}\,\xi\, .
\ee
This defines a cell, periodic rectangle, in plane with corners of the rectangle being at $0$, $4K$, $2iK'$ and $4K+2iK'$
which because of periodicity corresponds to a torus up to singularities.

\vspace{4em}
\begin{center}
\setlength{\unitlength}{.012in}
\begin{picture}(240,200)

\linethickness{0.075mm}
\put(0,103){\vector(1,0){230}}
\put(92,0){\vector(0,1){210}}
\put(40,20){\line(0,1){15}}
\put(142,20){\line(0,1){15}}
\put(192,20){\line(0,1){15}}

\put(40,45){\line(0,1){15}}
\put(142,45){\line(0,1){15}}
\put(192,45){\line(0,1){15}}

\put(40,70){\line(0,1){15}}
\put(142,70){\line(0,1){15}}
\put(192,70){\line(0,1){15}}

\put(40,95){\line(0,1){15}}
\put(142,95){\line(0,1){15}}
\put(192,95){\line(0,1){15}}

\put(40,120){\line(0,1){15}}
\put(142,120){\line(0,1){15}}
\put(192,120){\line(0,1){15}}

\put(40,145){\line(0,1){15}}
\put(142,145){\line(0,1){15}}
\put(192,145){\line(0,1){15}}

\put(40,170){\line(0,1){15}}
\put(142,170){\line(0,1){15}}
\put(192,170){\line(0,1){15}}


\put(21,28){\line(1,0){15}}
\put(21,66){\line(1,0){15}}
\put(21,141){\line(1,0){15}}
\put(21,178){\line(1,0){15}}

\put(46,28){\line(1,0){15}}
\put(46,66){\line(1,0){15}}
\put(46,141){\line(1,0){15}}
\put(46,178){\line(1,0){15}}

\put(71,28){\line(1,0){15}}
\put(71,66){\line(1,0){15}}
\put(71,141){\line(1,0){15}}
\put(71,178){\line(1,0){15}}

\put(96,28){\line(1,0){15}}
\put(96,66){\line(1,0){15}}
\put(96,141){\line(1,0){15}}
\put(96,178){\line(1,0){15}}
\put(121,28){\line(1,0){15}}
\put(121,66){\line(1,0){15}}
\put(121,141){\line(1,0){15}}
\put(121,178){\line(1,0){15}}

\put(146,28){\line(1,0){15}}
\put(146,66){\line(1,0){15}}
\put(146,141){\line(1,0){15}}
\put(146,178){\line(1,0){15}}

\put(171,28){\line(1,0){15}}
\put(171,66){\line(1,0){15}}
\put(171,141){\line(1,0){15}}
\put(171,178){\line(1,0){15}}

\put(196,28){\line(1,0){15}}
\put(196,66){\line(1,0){15}}
\put(196,141){\line(1,0){15}}
\put(196,178){\line(1,0){15}}

\thicklines
\put(92,103){\line(1,0){50}}
\put(92,102.5){\line(0,1){39}}
\put(92,141){\line(1,0){50}}
\put(142,102.5){\line(0,1){39}}
 \put(151,93){\makebox(0,0)[b]{\tiny{$4K$}}}
\put(105,144){\makebox(0,0)[b]{\tiny{$2iK'$}}}
 \put(239,101){\makebox(0,0)[b]{\tiny{$\xi_1$}}}
 \put(92,214){\makebox(0,0)[b]{\tiny{$\xi_2$}}}
\end{picture}
 \end{center}
\begin{center}
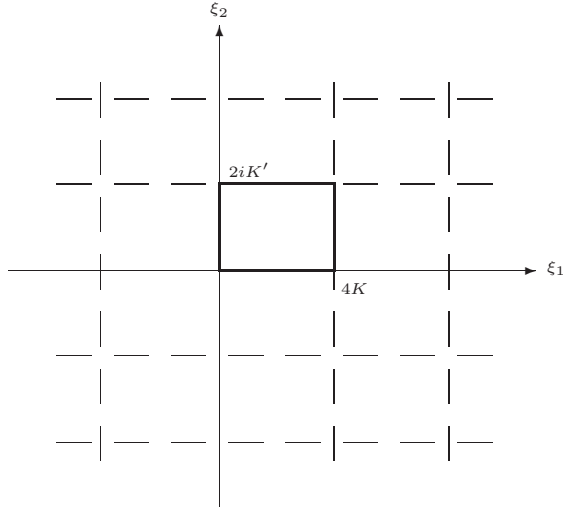
 Period Parallelograms for $\text{sn}\, \xi$ and $\xi=\xi_1+i \xi_2$.
\end{center}
 \vspace{1em}



To study the singularity structure of this solution, firstly, we consider the conformal-type singularities which are at branch points. This can be manifested by looking at the Einstein tensor,
\be
G_{z\bar z}=\frac{2}{\xi(z)+\xi(\bar z)}\left(\partial\,\frac{1}{\sqrt{(1-\bar z^2)(1-k^2\bar z^2)}}+\bar\partial\,\frac{1}{\sqrt{(1-z^2)(1-k^2z^2)}}\right)
\ee
By Stokes theorem \eqref{Stokes}, one should  compute the following contour integral
\be
\oint_{C}\frac{dz}{\sqrt{(1- z^2)(1-k^2 z^2)}}\,.
\ee
where $C$ is a large circle around origin. If we take the limit near the singular points we have,
\ba
\frac{1}{\sqrt{(1-z^2)(1-k^2z^2)}}&\simeq&\frac{1}{\sqrt{2}}\frac{1}{\sqrt{1-k^2}}\frac{1}{\sqrt{1\mp z}}\quad\text{as}\quad z\rightarrow \pm1\nonumber\\
&\simeq&\frac{1}{\sqrt{2}}\frac{k}{\sqrt{k^2-1}}\frac{1}{\sqrt{1\mp kz}}\quad\text{as}\quad z\rightarrow \pm 1/k\,.
\ea
Comparing with equation \eqref{delta1} this corresponds to $s=-1/2$ and consequently,
\be
G_{\theta\theta}=r^2G_{rr}=2r^2G_{z\bar z}=0,\qquad z\rightarrow \pm1,\pm1/k\,.
\ee
The contribution from $G_{\mu\nu}$ is then (for the conformal singularities only),
\ba
G_{\mu}^\nu&=&-\frac{8}{\tilde{\lambda} \ell^2}f\bar f\,\partial\bar \partial \ln f\bar f\,\delta_\mu^\nu\nonumber\\
&=&\frac{8}{\tilde{\lambda} \ell^2}f\bar f\left[\partial\left(\frac{\bar z}{1-\bar z^2}+\frac{k^2\bar z}{1-k^2\bar z^2}\right)+\bar\partial\left(\frac{z}{1-z^2}+\frac{k^2 z}{1-k^2z^2}\right)\right]\delta_\mu^\nu \, .
\ea
Using the Stokes theorem \eqref{Stokes}, we consider the following integral:
\ba
I&=&i\oint_C\left(\frac{\bar z }{1-\bar z^2}+\frac{k^2\bar z }{1-k^2\bar z^2}\right)d\bar z-\left(\frac{ z }{1- z^2}+\frac{k^2z }{1-k^2 z^2}\right)dz\nonumber\\
&=&-2\pi\int_Rd^2z \left[\delta^2(z-1,\bar z-1)+\delta^2(z+1,\bar z+1) \right.\nonumber\\
&&\qquad\qquad\qquad\left.+\delta^2(z-1/k,\bar z-1/k)+\delta^2(z+1/k,\bar z+1/k)\right]
\ea
 where $\partial R=C$. So we have,
\ba
G_{\mu}^\nu&=&-\frac{16\pi}{\tilde{\lambda} \ell^2}f\bar f\big[\delta^2(z-1,\bar z-1) +\delta^2(z+1,\bar z+1)\nonumber\\
&&\qquad\qquad +\delta^2(z-1/k,\bar z-1/k)+\delta^2(z+1/k,\bar z+1/k)\big]\delta_\mu^\nu \, .
\ea
Comparing with \eqref{Branecont}, we find the 3-brane tension at each branch point as,
\be
T_3=\frac{\pi}{\kappa^2}\,.
\ee
As expected this is the same as \eqref{T3-alpha} with $\alpha=2$.

On the other hand, although the internal metric as written in $z$ coordinate is periodic by default, the boundary conditions for the warp
factor should be imposed such that it is a periodic function in the $\xi$ plane. The warp factor is $W^2= (\xi+\bar\xi+c_1)^2$ on the fundamental
rectangle with $c_1$ an integration constant. This function can be extended periodically to the entire $\xi$-plane, however at $\xi_1=0, 4K$
boundaries it can not be pasted smoothly. We can choose $c_1=-4K$ such that $W^2= (2|\xi_1|-4K)^2$ to be periodic on the given lattice with discontinuous derivatives at $\xi_1=0, 4K$. This introduces warp-type singularities in the Einstein tensor and can be found as,
\ba
G_{\mu\nu}&=&-12\lambda \,K \,\delta(\xi_1)  \,\tilde g_{\mu\nu} \nonumber\\
G_{\xi_2\xi_2} &=& -\frac{4}{K}\, \delta(\xi_1) \nonumber\\
G_{\xi_1\xi_1} &=&0
\ea
To satisfy these matching conditions we need to smear 3-branes 
on 4-branes world volume \cite{Leblond:2001xr,Burgess:2001bn,Afshar:2009ps}.  Using \eqref{Branecont} we find,
\ba
T_{\mu\nu}&=&-\kappa^2\left(T_4+\frac{T_3}{L_{\xi_2}}\right)\lambda^{3/2}2 \,K^2  \,\delta(\xi_1) \,\tilde g_{\mu\nu}\nonumber\\
T_{\xi_2\xi_2} &=& -\kappa^2T_4\sqrt{\frac{\lambda}{4}} \,\delta(\xi_1) \nonumber\\
T_{\xi_1\xi_1} &=&0
\ea
where $L_{\xi_2}=K'\sqrt{\lambda}$ is the circumference of the 4-brane. As a result
\be
T_4=\frac{8}{\kappa^2K\sqrt{\lambda}}\qquad\text{and}\qquad T_3=-\frac{2K'}{\kappa^2K}.
\ee
Negative tension 3-branes can be interpreted as orientifold planes, O3-planes. Despite $T_3$ being negative, the
total effective tension $T_4 + T_3/L_{\xi_2}$ is positive.

The location of isolated 3-branes are at $\xi(\pm 1)=(2\mp 1)K$ and  $\xi(\pm 1/k)=(2\mp 1)K+i K'$ and the 4-brane is along the cycle $\xi_1=0$ as shown in the following figure.

\vspace{4em}
\begin{center}
\setlength{\unitlength}{.012in}
\begin{picture}(240,165)

\linethickness{0.075mm}
\put(0,50){\vector(1,0){230}}
\put(92,0){\vector(0,1){160}}

\put(92,50){\line(1,0){100}}
\put(92,125){\line(1,0){100}}

\linethickness{0.7mm}
\put(92,49.5){\line(0,1){75.5}}
\put(192,49.5){\line(0,1){75.5}}

\put(200,35){\makebox(0,0)[b]{$4K$}}
\put(110,131){\makebox(0,0)[b]{$2iK'$}}
 \put(239,47){\makebox(0,0)[b]{$\xi_1$}}
 \put(96,163){\makebox(0,0)[b]{$\xi_2$}}

\put(117,46.5){\makebox(0,0)[b]{$\bullet$}}
\put(117,83){\makebox(0,0)[b]{$\bullet$}}
\put(167,46.5){\makebox(0,0)[b]{$\bullet$}}
\put(167,83){\makebox(0,0)[b]{$\bullet$}}
\end{picture}
 \end{center}
\begin{center}
 Location of branes in $\xi$-plane.
\end{center}
 \vspace{1em}


Now we calculate the Euler character using \eqref{TrGauss} and \eqref{Euler}.  In this configuration there are boundaries but $\hat k=0$ so  as a result we have,
\ba\label{EULER}
\chi_{{}_E}=\frac{1}{2\pi}\int d^2y \,\sqrt{\hat g}\hat K(y)\,
&=&\frac{1}{2\pi}\int d^2y \,\sqrt{\hat g} \,\left[2e^{-w}\hat\nabla^2e^{w} +\kappa^2 \frac{7-p}{4}\,T_p\delta(\Sigma)\right]  \, . \nonumber\\
\ea
The contribution of the 4-brane  to the Euler character is,
\be
\chi^{(4)}_{{}_E}=\frac{\kappa^2}{2\pi}\int_0^{2K'}d\xi_2\int_0^{4K}d\xi_1\frac{3/4}{\kappa^2K}\delta(\xi_1)=\frac{3K'}{\pi K}
\ee
while the contributions of the 3-branes, both local and smeared ones, are
\ba
\chi^{(3)}_{{}_E}&=&\chi^{(3)}_{{}_E}|_\mathrm{isolated}+\chi^{(3)}_{{}_E}|_\mathrm{smeared}\nonumber\\
&=&2-\frac{\kappa^2}{2\pi}\int_0^{2K'}d\xi_2\int_0^{4K}d\xi_1\frac{2K'}{\kappa^2K}\delta(\xi_1)\delta(\xi_2)=2-\frac{K'}{\pi K} \, .
\ea
Finally the contribution from the warp factor in \eqref{EULER} gives
\be
\chi^{(w)}_{{}_E}=\frac{1}{2\pi}\int_0^{2K'}d\xi_2\int_0^{4K}d\xi_18 (2|\xi_1|-4K)^{-1}\delta(\xi_1)=-\frac{2K'}{\pi K}
\ee
Combining all, the Euler number of the internal space is,
\be
\chi_{{}_E}=\chi^{(w)}_{{}_E}+\chi^{(3)}_{{}_E}+\chi^{(4)}_{{}_E}=2\,.
\ee
This can be viewed as a pillow geometry with four corners. Note that the contributions of all warp-type singularities are canceled out while the contribution of the conformal-type singularities survives. This is expected since $\hat K$ is defined in terms of two-dimensional internal metric and thus depends on conformal singularities.

In order to compute the effective Planck mass we use the equation \eqref{eom1111} and \eqref{eom121},
\ba
M_{P}^2=\frac{\tilde{\lambda}^3 \ell^2}{8^3\kappa^2 }\int d^2z\,\partial W^2\bar \partial W^2
&=&\frac{\tilde{\lambda}^3 \ell^2}{8^3\kappa^2 }\int d^2z \,\left| 2W\partial W \right|^2    \nonumber\\
&=&\frac{\tilde{\lambda}^3 \ell^2}{8^3\kappa^2 }\int d^2z \,\left|2\frac{W}{f(z)}\right|^2    \nonumber\\
&=&\frac{\tilde{\lambda}^3 \ell^2}{128\kappa^2 }\int d^2\xi \,|\xi+\bar\xi|^2 \nonumber\\
&=&\frac{\tilde{\lambda}^3 \ell^2}{32\kappa^2 }\int_0^{2K'}d\xi_2 \int_0^{4K} d\xi_1 \,\xi_1^2 = \frac{4 \tilde{\lambda}^3 \ell^2 K' K^3}{3\kappa^2 }\,
\ea
and the result is finite.


Let us test the MN no-go theorem in \eqref{no-go} for this solution. Based on  the assumption in \eqref{EULER} which is the compactness
of the internal manifold, equation \eqref{no-go} leads to,
\ba\label{no-gotest}
\left(\frac{\tilde{\lambda}}{4}\right)^4 \ell^2 2K'\left[4^2\int_0^{4K} d\xi_1\, \xi_1^6-\left(\xi_1^4\partial_1\xi_1^4\right)_{\xi_1=4K}\right]-\kappa^2 T_4\int_Y(2|\xi_1|-2K)^6\delta(\xi_1)= \nonumber\\
2\cdotp4^4\tilde{\lambda}^4\ell^2 K^7K'\left(\frac{4}{7}-1\right)-\kappa^2 T_4 2 K' K^6 \leq0
\ea
so the MN no-go theorem is bypassed as expected.

\section{Solutions with $\boldsymbol{\sigma =1}$}
\label{nonflat}

Now we consider the general case  $\sigma \neq 0$ in \eqref{eom121} which as mentioned before is equivalent to $\sigma=1$.
Here we are interested in the warp factor which depends only on the radial direction $r=\sqrt{z\bar z}$. The only way to get this ansatz is via $f= z$.
 Then equations \eqref{eom121} merge to a single form
\be\label{443}
\frac{dW}{d\rho}=2(1+\frac{1}{W^3}) \, ,
\ee
where $\rho=\ln r$. This can be solved by
\be\label{eom333}
W-\frac{1}{6}\left[\ln \left( \frac{\left( W+1 \right)^2}{ \left( W-1 \right)^2+W}\right) +2\sqrt
{3}\arctan \left( \,{\frac {  2\,{W}-1  }{\sqrt {3}}} \right) \right]=2\ln \frac{r}{r_0}\,.
\ee
The line element becomes
\ba
ds^2=\frac{\tilde{\lambda}^2}{16}W^2\tilde g^{\text{dS}}_{\mu\nu}dx^\mu dx^\nu +\ell^2\frac{\tilde{\lambda} }{8}
\partial_\rho W\left(d\rho^2+d\theta^2\right)\, .
\ea
 The implicit solution of $W$ in \eqref{eom333} indicates that it is difficult to handle $W$ as a function of $\rho$, however we may interchange roles of $W$ and $\rho$ by considering \eqref{443} as a coordinate transformation to find the following metric in which $W$ is treated as a coordinate:
\ba
ds^2=\frac{\tilde{\lambda}^2}{16}W^2\tilde g^{\text{dS}}_{\mu\nu}dx^\mu dx^\nu + \ell^2
\frac{\tilde{\lambda}}{4}\left(\frac{dW^2}{4(1+W^{-3})}+(1+W^{-3})d\theta^2\right)\,.
\ea
 This is a Ricci flat solution by construction, however, it is singular at $W=0$, since:
\ba
R_{\mu\nu\rho\sigma}R^{\mu\nu\rho\sigma}=\frac{61440}{\tilde{\lambda}^2 W^{10}} \, .
\ea
To avoid this singularity, we can cut a slice of $W$ in which $(1+W^{-3})\geq 0$ and thus $\lambda>0$. Moreover, we need to choose this slice and other free parameters such that at the end we get positive tension branes. This can be achieved by the following coordinate transformation,
\ba
W(u)=-|u|^3-u^2+|u|-\sqrt[3]{2}
\ea
where $|u|\leq 1/3$. This corresponds to the identification of the positive and negative values of $u$ with periodic smooth boundary condition at $u=\pm 1/3$ where $\partial_u W=0$ and the only singularity is at $u=0$. By the above transformation we find the following metric
\ba
ds^2 &=&\frac{\tilde{\lambda}^2}{16}W(u)^2\tilde g^{\text{dS}}_{\mu\nu}dx^\mu dx^\nu \nonumber\\
&&+\ell^2\frac{\tilde{\lambda}}{4}\left(\frac{ (-3u^2-2|u|+1)^2 du^2} {4(1+W(u)^{-3})}+(1+W^{-3})d\theta^2\right)\,.
\ea
Regarding the singularity at $u=0$, it should be accompanied by a brane at $u=0$. To find the brane tension, we use
\ba
G^\mu_\nu&=&-\frac{36 \times 2^{2/3}}{\ell^2\tilde{\lambda}}\delta(u) \delta^\mu_\nu \nonumber\\
G^\theta_\theta &=& -\frac{32 \times 2^{2/3}}{\ell^2\tilde{\lambda}}\delta(u) \nonumber\\
G^u_u &=&0
\ea
This corresponds to a singularity sourced by 3-branes in $\mu\nu$ directions
smeared over a 4-brane in $\mu\nu$ and $\theta\theta$ directions \cite{Leblond:2001xr,Burgess:2001bn,Afshar:2009ps}.

Using \eqref{Branecont} we obtain
\ba
\mu\nu:\;\;\; T_4+\frac{T_3}{L_\theta} &=& \frac{36 \times 2^{2/3}}{\kappa^2\sqrt{\tilde{\lambda}}} \nonumber\\
\theta\theta:\;\;\; T_4 &=& \frac{32 \times 2^{2/3}}{\kappa^2\sqrt{\tilde{\lambda}}}
\ea
In which $L_\theta$ is the circumference of the 4-brane given by $L_\theta=\pi \ell\sqrt{\tilde{\lambda}}(1+W(0)^{-3})^{1/2}$. As a result we find
\ba
T_3= \frac{4\pi \times 2^{1/6}}{\kappa^2 }
\ea
Interestingly, both 3- and 4-brane tensions are positive.

This solution has two compact coordinates as $\theta$ and $u$ where both are periodic. So we expect  it to be a torus with two cycles around $\theta$ and $u$.


The 4-D Planck mass is obtained to be
\ba
M_{P}^2
&=& \kappa^{-2}\int d^2z \sqrt{\hat g}e^{2w} \nonumber\\
&=& \frac{\tilde{\lambda}^3\ell^2}{128  \kappa^2}\int_{-1/3}^{1/3} W(u)^{2}(-3u^2-2|u|+1)du \nonumber\\
&=& \frac{5 \left(25-405\times 2^{1/3}+2187\times 2^{2/3}\right) }{3779136}\frac{\tilde{\lambda}^3\ell^2}{\kappa^2}
\simeq  0.004 \frac{\tilde{\lambda}^3\ell^2}{ \kappa^2}
\ea
which is finite.

In order to generalize the above case let us assume that  the warp factor depends only on $\sqrt{Z(z)\bar Z(\bar z)}=|Z(z)|$ where $Z(z)$ is an arbitrary holomorphic function. Then  equations \eqref{eom121} become
\ba\label{eom128}
\frac{fZ'}{2}\sqrt{\frac{\bar Z}{Z}}\frac{d W}{d|Z|}=1+\frac{1}{W^3}\quad\text{and}\quad \frac{\bar f \bar Z'}{2}\sqrt{\frac{ Z}{\bar Z}}\frac{d W}{d|Z|}=1+\frac{1}{W^3}\,.
\ea
By taking $f=Z/Z'$, these equations combine  into a single equation
\be
\frac{d W}{d(\ln|Z|)}=2(1+\frac{1}{W^3})\,
\ee
and the same story applies here as in \eqref{443}.

\section{Summary and conclusions}\label{summary}

In this paper we studied the phenomenological solutions for the six-dimensional gravitational set up  with axion-dilaton and local branes in the context of brane-world scenario. We have found two different classes of
maximally symmetric compactifications. The first class consists of a flat four-dimensional space with no warp factor with axion-dilaton field,
$\tau=\tau(z)$, as a holomorphic function of two-dimensional extra manifold.  This indeed has already been  studied in the context of F-theory compactification of
IIB supergravity to eight dimensions with the space-filling  D7-branes \cite{Greene:1989ya,   Bergshoeff:2006jj}. There, using SL$(2,\mathbb{R})$ invariance of the solution, one finds 3-branes
as isolated conical singularities in the complex plane with tensions given in terms of deficit angles.

In section \ref{consttau}, we introduced our second class of solutions with constant $\tau$, i.e. pure gravity, in which the  compactification yields four-dimensional dS space ($\lambda>0$) with non-trivial warp factor. Different solutions are characterized by  an arbitrary holomorphic function $f$. Locally different choices of $f$ are equivalent but globally they differ by  singularities arising from branes positions.
We have considered different classes of singularities for $f$. The first example corresponds to a cylindrical configuration with $f$ being an exponential function of the complex $z$ coordinate. The geometry is cut at $z=0$ and the two pieces are identified under the $\mathbb Z_2$ symmetry.
This choice includes a 4-brane at the boundary with a positive tension, scaling like
$1/\sqrt{\lambda}$. It also gives a finite four-dimensional gravity.

As an another interesting choice we have considered $ f \sim z^{s}$ with $s$ a half-integer. This involves branch cut starting at $z=0$ which gives conical singularity
with deficit angle $-2\pi s$, so a 3-brane is placed at $z=0$. The physically allowed configuration corresponds to $s= -\frac{1}{2}$ with the deficit angle $\pi$.
To find finite four-dimensional gravity, one may cut the space by a 4-brane to
find a disc (a finite cone). This is the inversion of cylindrical solution which corresponds to the outer region of the disc. In both cylindrical
and conical solutions, the warp factor is a function of both radial and azimuthal coordinates. Interestingly, the solution is single-valued and
smooth when going around the azimuthal direction.

In our third example a multiple branch point solution was introduced for which the warp factor can be written in terms of an inverse Jacobian elliptic function.
It leads to a pillow topology with four 3-branes at branch points with positive tensions and a 4-brane wrapped over a cycle of torus with 3-branes wrapped on its world volume. Despite the smeard 3-brane tension being negative, the
total effective tension is positive. The four-dimensional Planck mass is finite too.

In the first two cases we need to consider 4-branes to find a finite 4D Planck
mass, in the third example the finiteness of the effective 4D Planck mass is
fulfilled by using a double periodic function and consequently restricting
the limits of the 2D integrals to its periodicity. But then to make the
warp factor also periodic we again have to introduce junctions which
amounts to using 4-branes. We could not find a holomorphic function that
does not need the introduction of 4-branes.

In these solutions, by bypassing the Maldacena-Nunez theorem, we have shown that for pure 6-dimensional gravity, dS solutions with non-trivial warp factor and finite four-dimensional  gravity are possible. The solutions involve  three- and four-branes with positive/negative tensions 
in which branch cuts and points may appear depending on the configurations.

In our setup, we distinguished two types of singularities and called them warp-type and conformal-type singularities, as they were originated, respectively, from four-dimensional warp factor and two-dimensional conformal factor in the metric. This classification especially helps us when studying the pillow-like solution mentioned above. The conformal-type singularities lead to conical 3-branes and warp-type lead to 4-brane (and smeared 3-branes) wrapped around a non-trivial cycle.

The global topology of solutions was examined by calculating the Euler character. In the case of pillow geometry, it was shown that only the conical three-branes contribute to the Euler number. This can be explained by calculating the Euler number directly from the internal metric in which it manifestly depends on conformal factor in the metric and as a result it depends on the conformal-type singularities or conical 3-brane. On the other hand the calculation of the  Euler character by using the equations of motion, as we did here, involves the warp factor. Nonetheless, the singularity from the warp factor was consistently canceled by the contribution from the four-brane and smeared three-branes.

There are different phenomenological directions which this setup can be used.
One application is towards the brane world scenario which is studied extensively in the past, 
for example see  \cite{Randall:1999ee, Aghababaie:2003ar}. In this view the Standard Model of particle physics can be localized in each of localized sources in this set up. Depending on the warp factor, the four-dimensional physical mass scale can have different values which may be used to explain the hierarchy problem. 
As for other application, one can look into embedding brane inflation in this set up, following the idea in  \cite{Dvali:1998pa, Kachru:2003sx}. In this view, one can put a pair of brane and anti-brane in this setup and look for the inflationary potential. The advantage in this set up is that  in 6D we have a better control on the volume modulus so in principle the back-reactions of
volume modulus on inflaton field is better understood. This is in contrast to usual brane inflation 
in ten-dimensional compactifications in which due to global and local effects, the back-reaction effects from the Kahler modulus field is not under control. 

One shortcoming of this work is that no stability analysis is performed. As is typical in many
phenomenological brane world scenarios, usually the volume modulus is not stable, as an example see \cite{Burgess:2001bn}. It is a non-trivial question whether or not our set up is stable under perturbations. This is an open question which is beyond the scope of this work.

\acknowledgments{
We are grateful to Andreas Braun, Cliff Burgess, Daniel Grumiller, Anshuman Maharana
and Fernando Quevedo for useful discussions and comments on the draft. H.A. and H.F. would like to thank
the Isaac Newton Institute for Mathematical Sciences for the hospitality during the program ``Mathematics
and Applications of Branes in String and M-theory'' when this work was in progress. H.A. also thanks
the institute for research in fundamental sciences, IPM, for hospitality at different stages of this work.
H.A. was supported by the projects I 952-N16, Y 435-N16 and P 21927-N16 of the Austrian Science Fund (FWF).}


\appendix


\section{Brane contribution}
\label{brane-T}
Here we briefly outline the contributions of local D$_p$-branes into the energy momentum tensor.

For a D$_p$-brane wrapping a ($p-3$)-cycle $\Sigma$ in the $D$-dimensional space-time $M$ the relevant interactions are
\be
S_{\text{loc}}=-\int_{\mathbb{R}^4\times\Sigma}d^{p+1}\xi\,  T_p\sqrt{-|^{(p+1)}g|}+\mu_p\int_{\mathbb{R}^4\times\Sigma}C_{p+1}\,.
\ee
where $|^{(p+1)}g|$ is the determinant of the brane world volume metric. Branes which are only gravitationally coupled have $\mu_p=0$ and may in general be called $p$-branes. 

The energy-momentum tensor associated with these local sources are given by
\ba
T^{\text{loc}}_{MN}=-\frac{2}{\sqrt{-g}}\frac{\delta S_{\text{loc}}}{\delta g^{MN}} \, .
\ea
Here $S_{\text{loc}}$ is the action describing the sources. We find $T^{\text{loc}}_{MN}$ as
\be\label{Branecont}
T^{\text{loc}}_{MN}=-T_p\; {^{(p+1)}g_{IJ}}\delta^{I}_M\delta^{J}_N\,\delta(\Sigma)
\ee
where
\be
\delta(\Sigma)=\sqrt{\frac{^{(p+1)}g}{g}}\;\delta^{D-p-1}(y-y_0) \, ,
\ee
and $y_0$ is the position of the brane. It is then easy to show that
\be
(T^{\mu}_{\mu})_{\text{loc}}=-4\,T_p\delta(\Sigma)\qquad\text{and}\qquad (T^m_m)_{\text{loc}}=-(p-3)\,T_p\delta(\Sigma)\,.
\ee
Using the trace-reversed form of Einstein equation,
\be
R_{MN}=\kappa^2\left(T_{MN}-\frac{1}{D-2}g_{MN}T^L_L\right)
\ee
we find,
\ba\label{Branecont1}
(g^{\mu\nu}R_{\mu\nu})_{\text{loc}}&=&4\kappa^2\left( \frac{3+p-D}{D-2} \right)T_p\delta(\Sigma)\\
(g^{mn}R_{mn})_{\text{loc}}&=&2\kappa^2\left( \frac{2D-p-5}{D-2} \right)T_p\delta(\Sigma) \, .
\ea
\section{Singularity analysis}
\label{singular}
Here we present some local singularity analysis in the complex plane which will be useful
in determining the local 3-branes tensions. To fix the notation we consider the two-dimensional manifold $Y$ with the follwing  metric,
\be
\label{2dmetr}
d{\hat s}^2= e^{\Omega(r,\theta)} d{\bar s}^2=
e^{\Omega(r,\theta)} (dr^2+r^2 d\theta^2)=e^{\Omega(z,\bar z)}dzd\bar z
\ee
We have
\be\label{gaussian}
2\hat K(y)=e^{-\Omega}(2\bar K(y)-\bar
\nabla^2\Omega)\quad\text{and}\quad\bar\nabla^2\Omega=-4\partial\bar\partial\Omega\,,
\ee
where $\hat K(y)$ is the gaussian curvature of $Y$ and bared quatities are w.r.t. the flat metric.

For a general vector in the complex plane $v=v^z\partial+v^{\bar z}\bar \partial$ whose components $v^z$ and $v^{\bar z}$ are analytic ((anti-)holomorphic)
functions in a region $R$ and its boundary $C$, we have the following Stokes theorem,
\be\label{Stokes}
\int_Rd^2z\left(\partial v^z+\bar \partial v^{\bar z}\right)=i\oint_{\partial R}\left( v^zd\bar z-v^{\bar z}d z\right)\,.
\ee
Defining\footnote{$\delta^2(z,\bar z)=\frac{1}{2}\delta(y_1)\delta(y_2)$}
\be
\int d^2z \,\delta^2(z,\bar z)=1
\ee
we find
\be\label{delta}
\partial\frac{1}{\bar z}+\bar \partial\frac{1}{ z}=4\pi \delta^2(z,\bar z)\, ,
\ee
where we have used the fact that
\ba\label{residue}
\oint_Cz^ndz= \left\{
  \begin{array}{cc}
     0 &\; n\neq-1  \nonumber\\
    2\pi i &\; n=-1  \nonumber\\
  \end{array}
\right.\qquad\text{and}\qquad n\in \mathbb{Z}
\ea
where $C$ is the unit circle surrounding the origin. How about non-integer powers? We know that $z^s$ is in general  a multi-valued function and is not well-defined
unless $s\in \mathbb{Z}$ or including a branch cut which limits the domain of the function to a specific branch. In order to solve this integral in the presence of the branch cut
 we setup a closed contour, $\Gamma$, which does not cross the branch cut and surrounds the origin. This contour consists of four pieces; the unit circle $C$, a
line above the branch cut towards the origin, $L_1$, a circle of radius $\epsilon$ around the origin, $C_\epsilon$,  and again a line $L_2$ under the branch cut which ends back to
$C$
\be
\oint_\Gamma=\int_{C}+\int_{L_1}+\int_{C_\epsilon}+\int_{L_2} \, .
\ee
The integral around $\Gamma$ vanishes because it neither surround any pole nor crosse any branch cut. In the limit $\epsilon \rightarrow0$ the integral
around $C_\epsilon$ becomes zero when $s>-1$,
\be
\oint z^s dz=i\epsilon^{s+1}\int e^{i\theta(s+1)}d\theta\,\longrightarrow0 \quad\text{as}\quad\epsilon\rightarrow0\,.
\ee
The integral along $L_1$ and $L_2$ differ by a phase $e^{2\pi i s}$, so we have
\ba
\oint_Cz^sdz= \frac{1}{s+1}\left(e^{2\pi i s}-1\right) \, .
\ea
Note that this result is true even for integer values of $s>-1$, for $s=-1$ one should take the limit $s\rightarrow-1$ to find \eqref{residue}.
Applying the Stokes theorem \eqref{Stokes} we find
\be\label{delta1}
{ \partial \bar z^s+\bar \partial z^s=\frac{2\sin 2\pi s}{s+1} \,\delta^2(z,\bar z)\qquad\text{for}\qquad s>-1.}
\ee
This is obviously zero for all integer values of $s$ as expected, it also gives the correct result in \eqref{delta} when taking the limit $s\rightarrow-1$.
There are however non-integer values of $s$ for which the delta
function singularity in \eqref{delta1} also goes away:
\ba\label{svalues}
s&=&k+\frac{1}{2}\qquad\text{and}\quad k\in\mathbb{Z}\\
&=&\pm\frac{1}{2},\,\frac{3}{2},\,\frac{5}{2},\,\cdots\, ,
\ea
which are the only compatible ones with the condition $s>-1$. This is interesting because had not we added the contribution from
the holomorphic part with the contributions from the antiholomorphic part, we could not have seen it.


\providecommand{\href}[2]{#2}\begingroup\raggedright\endgroup

\end{document}